\documentclass[12pt,aps,epsfig,amssymb,tighten]{revtex4}
\usepackage{epsfig}
\begin{document}
\title{Equilibrium of a confined, randomly-accelerated,\\
inelastic particle: Is there inelastic collapse?}
\author{
Theodore W. Burkhardt and Stanislav N. Kotsev}
\affiliation{Department of Physics, Temple University,
Philadelphia, PA 19122}
\date{\today}

\begin{abstract}
We consider the one-dimensional motion of a particle randomly
accelerated by Gaussian white noise on the line segment $0<x<1$.
The reflections of the particle from the boundaries at $x=0$ and 1
are inelastic, with velocities just after and before reflection
related by $v_f=-rv_i$. Cornell {\it et al.} have predicted that
the particle undergoes inelastic collapse for $r<r_c=e^{-\pi/\sqrt
3}=0.163$, coming to rest at the boundary after an infinite number
of collisions in a finite time and remaining there. This has been
questioned by Florencio {\it et al.} and Anton on the basis of
simulations. We have solved the Fokker-Planck equation satisfied
by the equilibrium distribution function $P(x,v)$ with a
combination of exact analytical and numerical methods. Throughout
the interval $0<r<1$, $P(x,v)$ remains extended, as opposed to
collapsed. There is no transition in which $P(x,v)$ collapses onto
the boundaries. However, for $r<r_c$ the equilibrium boundary
collision rate is infinite, as predicted by Cornell {\it et al.},
and all moments $\overline{\thinspace|v|^q}$, $q>0$ of the
velocity just after reflection from the boundary vanish.
\end{abstract}
\maketitle

\newpage
\section{Introduction}
\label{sec:intro} Consider a particle randomly accelerated on the
line segment $0<x<1$ according to
\begin{equation}{d^2x\over
dt^2}=\eta (t)\thinspace,\quad\langle\eta(t)\eta(t')\rangle=
2\delta(t-t')\thinspace,\label{eqmo}
\end{equation}
where $\eta(t)$ is uncorrelated white noise with zero mean. If the
collisions of the particle with the boundaries at $x=0$ and 1 are
elastic, the mean square velocity increases according to
\begin{equation}
\langle v(t)^2\rangle=\langle
v(0)^2\rangle+2t\thinspace,\label{ke}
\end{equation}
just as in the absence of boundaries.

In this paper we assume that the boundary collisions of the
randomly accelerated particle are {\em inelastic}. The velocities
just after and before reflection satisfy
\begin{equation}
v_f=-rv_i\thinspace,\label{restitution}
\end{equation}
where $r$ is the coefficient of restitution. This simple model is
of interest in connection with the statistics of driven granular
media, where particles tend to cluster, due to inelastic
collisions, even though no attractive forces are present. The
model was studied by Cornell, Swift, and Bray (CSB) \cite{csb},
who argued that the particle undergoes "inelastic collapse," i.e.
makes an infinite number of collisions in a finite time, comes to
rest at the boundary, and remains there, if the coefficient of
restitution $r$ is less than the critical value
\begin{equation}
r_c=e^{-\pi/\sqrt 3}=0.163\dots\label{rc}
\end{equation}

The prediction of inelastic collapse was questioned by Florencio
{\it et al.} \cite{fbb}, who carried out simulations and found
that the particle did not adhere to the boundary for any $r$.
Anton \cite{la} reported that his simulations are consistent with
an infinite collision rate for $r<r_c$ but also incompatible with
localization of the particle at the boundary.

According to Eqs. (\ref{ke}) and (\ref{restitution}) the kinetic
energy of the randomly accelerated particle increases in between
boundary collisions but decreases, for $r<1$,  in the collisions.
Eventually an equilibrium is reached. Burkhardt, Franklin and
Gawronski (BFG) \cite{bfg} analyzed the equilibrium distribution
$P(x,v)$ for the position and velocity of the particle for
$r_c<r<1$. This function satisfies the steady-state Fokker-Planck
equation
\begin{equation}
\left(v{\partial\over\partial x}-{\partial^2\over \partial
v^2}\right)P(x,v)=0\thinspace,\label{fp}
\end{equation}
with the boundary conditions
\begin{eqnarray}
P(x,v)&=&P(1-x,-v)\thinspace,\label{refsym}\\
P(0,-v)&=&r^2P(0,rv)\thinspace,\quad v>0\thinspace,\label{probcon}
\end{eqnarray}
corresponding to reflection symmetry and conservation of
probability, respectively. In particular, the second boundary
condition insures that the incident and reflected probability
currents at the boundary have equal magnitude
\begin{equation} I= \int_0^\infty
dv\thinspace vP(0,-v)=\int_0^\infty dv\thinspace
vP(0,v)\thinspace.\label{collrate}
\end{equation}

Making use of an exact Green's function solution of Eqs.
(\ref{fp})-(\ref{probcon}), BFG found that the boundary collision
rate $I$, defined by Eq. (\ref{collrate}), diverges as $r$
approaches $r_c$ from above and that $P(x,v)$ is extended, as
opposed to collapsed, at $r=r_c$. In this approach $P(x,v)$ is
obtained as the difference of two integrals, both of which diverge
for $r\leq r_c$. This was noted by BFG, who, however, incorrectly
concluded that the solution to the Fokker-Planck equation breaks
down for $r<r_c$.

In this paper the calculation of $P(x,v)$ is extended to $r<r_c$.
In Section II the approach of BFG is reviewed. The divergences,
for $r\leq r_c$, of the two integrals which determine $P(x,v)$ are
shown to cancel, leaving a finite result. Throughout the entire
interval $0<r<1$, $P(x,v)$ varies smoothly and analytically with
$r$. There is no transition in which $P(x,v)$ collapses onto the
boundaries.

In Section III the equilibrium boundary collision rate is
calculated from the results of Section II. The collision rate is
finite for $r>r_c$ and infinite for $r\leq r_c$, as predicted by
CSB. All the equilibrium moments $\overline{\thinspace|v|^q}$,
$q>0$ of the velocity just after striking the boundary
\cite{overline} vanish for $r\leq r_c$.

Our conclusions are summarized in Section IV, and some earlier
results on inelastic collapse are reexamined.

\section{Solution of the Fokker-Planck equation}
We begin with a brief review of the approach \cite{bfg} of BFG.
Generalizing earlier work of Masoliver and Porr\`a \cite{mp}, they
showed that the Fokker-Planck equation (\ref{fp}), with reflection
symmetry (\ref{refsym}), has the exact solution
\begin{equation}
P(x,v)=\int_0^\infty du\thinspace u\thinspace
G(x,v,u)P(0,u)\label{gf1}
\end{equation}
for $v>0$ in terms of the Green's function
\begin{eqnarray}
&&G(x,v,u)={v^{1/2}u^{1/2}\over 3x}\thinspace
e^{-(v^3+u^3)/9x}I_{-1/3}\left({2v^{3/2}u^{3/2}\over
9x}\right)\nonumber\\ &&\qquad\qquad -{1\over
3^{1/3}\Gamma({2\over 3})}\int_0^x dy\thinspace
{e^{-v^3/9(x-y)}\over
(x-y)^{2/3}}\left[R(y,u)-R(1-y,u)\right]\thinspace,\label{gf2}
\end{eqnarray}
where
\begin{equation}
R(y,u)={1\over 3^{5/6}\Gamma({1\over 3})\Gamma({5\over
6})}\thinspace {u^{1/2}e^{-u^3/9y}\over
y^{7/6}(1-y)^{1/6}}\thinspace _1F_1 \left(-{1\over 6},{5\over
6},{u^3(1-y)\over 9y}\right)\thinspace,\label{gf3}
\end{equation}
and $_1F_1(a,b,c)$ is a standard confluent hypergeometric function
\cite{gr}.

To calculate $P(x,v)$ from Eq. (\ref{gf1}), one must first
determine the unknown function $P(0,u)$ on the right hand side.
Setting $x=1$ in Eqs. (\ref{gf1})-(\ref{gf3}) and using
$r^2P(0,rv)=P(1,v)$, as follows from Eqs. (\ref{refsym}) and
(\ref{probcon}), leads to the integral equation
\begin{equation}
r^2P(0,rv)=\int_0^\infty du\thinspace u\thinspace
G(1,v,u)P(0,u)\thinspace\label{ieq1}
\end{equation}
for $P(0,v)$, where
\begin{equation}
G(1,v,u)={1\over 6\pi}v^{1/2}u^{1/2}\thinspace
e^{-(v^3+u^3)/9}\left[{9\over v^3+u^3}+6\thinspace
_1F_2\left(1;{5\over 6},{7\over 6};{v^3u^3\over
81}\right)\right]\thinspace,\label{ieq2}
\end{equation}
and $_1F_2(a;b,c;z)$ is a generalized hypergeometric function
\cite{gr}. The quantity $vG(1,v,u)$ is of interest in its own
right. As discussed in the Appendix, it generalizes McKean's
result \cite{mck} for the velocity distribution at first return to
the boundary from the half line $x>0$ to the line segment $0<x<1$.

BFG showed \cite{bfg} that the asymptotic form of $P(0,v)$ for
small and large $v$ is determined by the first and second terms,
respectively,  of the kernel $G(1,v,u)$ in Eqs. (\ref{ieq1}) and
(\ref{ieq2}) and given by
\begin{equation}
P(0,v)\sim\left\{\begin{array}{l}v^{-\beta(r)}\thinspace,\\e^{-v^3/v_{\rm
ch}(r)^3}\thinspace,\end{array}\right.\quad\begin{array}{l}v\to 0\thinspace,\\
v\to\infty\thinspace,\end{array}\label{asy1}
\end{equation}
where
\begin{eqnarray}
&&r=\left[2\sin\left({2\beta +1\over
6}\thinspace\pi\right)\right]^{1/(\beta
-2)}\thinspace,\label{asy2}\\&&v_{\rm ch}(r)^3={9r^3\over
1-r^3}\thinspace.\label{asy3}
\end{eqnarray}
Note the non-Maxwellian velocity distribution. As $r$ decreases,
the boundary collisions become more inelastic, and the probability
of finding the particle near the boundary with a small velocity
increases. This is seen in the monotonic increase of the exponent
$\beta(r)$ from 0 to ${5\over 2}$, as $r$ decreases from 1 to 0.
The characteristic velocity $v_{\rm ch}(r)$ also decreases with
decreasing $r$.

The asymptotic forms (\ref{asy1})-(\ref{asy3}) are smooth analytic
functions of $r$ throughout the interval $0<r<1$. There is no
singular behavior at $r_c$. In particular, on expanding the right
side of Eq. (\ref{asy2}) about $\beta=2$, one sees that $\beta(r)$
is a nonsingular function of $r$, with $\beta(r_c)=2$. To connect
the asymptotic forms (\ref{asy1})-(\ref{asy3}) of $P(0,v)$ for
small and large $v$, we have solved integral equation (\ref{ieq1})
by numerical iteration, as in \cite{bfg}. As noted by BFG
\cite{bfg}, the integral equation appears to have a well-defined
solution for $0<r<1$, i.e. $0<\beta<5/2$, with no special behavior
at $r_c$. Numerical results for several values of $r$ above and
below $r_c=0.163$, are shown in Fig. 1. The slopes of the curves,
for small $v$, depend on $r$ in accordance with the asymptotic
form (\ref{asy1}), (\ref{asy2}). There is no qualitative
difference above and below $r_c$. Presumably, $P(0,v)$, like its
exact asymptotic forms (\ref{asy1})-(\ref{asy3}) for small and
large $v$, is an analytic function of $r$ throughout the interval
$0<r<1$.

Once $P(0,v)$ has been determined, $P(x,v)$ may be obtained by
integration. According to Eqs. (\ref{gf1})-(\ref{gf3}), $P(x,v)$
is the sum of two integrals over $P(0,u)$, corresponding to the
two terms in the Green's function $G(x,v,u)$ in Eq. (\ref{gf2}).
Both integrals diverge at the lower limit for $r\leq r_c$, as
follows from the asymptotic form (\ref{asy1}), (\ref{asy2}) of
$P(0,u)$ for small $u$, with $\beta(r)>2$ for $r<r_c$. This was
noted by BFG \cite{bfg}, who, however, incorrectly concluded that
the solution to the Fokker-Planck equation breaks down for
$r<r_c$. The divergences cancel, leaving a finite result, as may
be seen by integrating with a low $u$ cutoff and sending the
cutoff to zero after adding the two integrals. No cutoff is needed
if the two integrands are added before integrating over $u$. From
Eq. (\ref{gf2}) it is straightforward to show that $G(x,v,u)\sim
u^{1/2}$ in the small $u$ limit \cite{smallu}. Thus the integral
in Eq. (\ref{gf1}) behaves as $\int_0 du\thinspace u^{3/2-\beta}$
for small $u$. Since $0<\beta(r)<{5\over 2}$ for ${0<r<1}$, there
are no convergence problems at the lower limit of the integral.
Throughout the interval $0<r<1$, $P(x,v)$ is a smooth well-defined
function of $r$, presumably analytic in $r$, and does not collapse
onto the boundaries at $x=0$ and $x=1$.

We have also considered the probability density
\begin{equation}
P(x)=\int_{-\infty}^\infty dv\thinspace P(x,v)=\int_0^\infty
dv\thinspace\left[P(x,v)+P(1-x,v)\right]\label{posprob1}
\end{equation}
for the position of the particle. From Eqs.
(\ref{gf1})-(\ref{gf3}) and (\ref{posprob1}), one finds that the
leading singular contribution to $P(x)$ for $x\to 0$ is determined
by the asymptotic form $P(0,v)\approx Av^{-\beta}$ for $v\to 0$ in
Eq. (\ref{asy1}) and given by
\begin{eqnarray}
&&P_{\rm sing}(x)\approx Bx^{(1-\beta)/3}\thinspace,\quad x\to 0\thinspace,\label{posprob2}\\
&&B={2\pi\over 3^{(4\beta +5)/6}}\thinspace{\Gamma({\beta-1\over
3})\over\sin\left({2\beta+1\over
6}\thinspace\pi\right)\Gamma({\beta\over 3})\Gamma({\beta+1\over
3})}\thinspace A\thinspace.\label{posprob3}
\end{eqnarray}
For $0<\beta<1$, i.e. ${1\over 2}<r<1$, the leading singular
contribution to $P(x)=P(1-x)$ in Eq. (\ref{posprob2}) vanishes as
$x$ approaches 0 or 1, and $P(0)$ is finite and nonzero. For
$\beta>1$ or $r<{1\over 2}$, $P(x)$ diverges according to Eq.
(\ref{posprob2}) as $x$ approaches 0 or 1. Since the divergence is
integrable, $P(x,v)$ can be normalized so that
\begin{equation}
\int_{-\infty}^\infty dv\int_0^1 dx\thinspace P(x,v)=\int_0^1
dx\thinspace P(x)=1\thinspace\label{normP}
\end{equation}
for all $0<r<1$. Note the absence of any special behavior in Eqs.
(\ref{posprob2}), (\ref{posprob3}) at $\beta=2$ or $r=r_c$.

The probability density $P(x)$ is shown for several values of $r$
above and below $r_c$ in Fig. 2. The curves were obtained by
integrating Eq. (\ref{gf1}) over $v$ analytically and then
performing the $u$ integration numerically, using the numerical
solution for $P(0,u)$ in Fig. 1.  Again there is no qualitative
difference above and below $r_c$.

\section{Collision Rate and Moments of the Reflected Velocity}

Unlike the distribution functions $P(x,v)$ and $P(x)$ considered
thus far, the equilibrium collision rate $I$, defined by Eq.
(\ref{collrate}), does indeed change non-analytically as $r$
passes through $r_c$. According to the asymptotic forms
(\ref{asy1})-(\ref{asy3}) of $P(0,v)$, the second integral on the
right of Eq. (\ref{collrate}) converges at the upper limit for all
$0<r<1$ and at the lower limit for $\beta<2$ but not $\beta\geq
2$. Thus the boundary collision rate is finite for $r>r_c$ and
infinite for $r\leq r_c$, in agreement with the prediction of CSB
\cite{csb}.

The moments $\overline{\thinspace |v|^q}$ of the velocity just
after reflection from the boundary \cite{overline} exhibit a
closely related collapse transition. Since $v P(0,v)dv$ is the
reflected probability current in the velocity range $v$ to $v+dv$,
\begin{equation}
\overline{\thinspace |v|^q}={\int_0^\infty dv\thinspace v^{q+1}
P(0,v)\over\int_0^\infty dv\thinspace vP(0,v)}\thinspace,\quad
r>r_c\thinspace.\label{A1}
\end{equation}
The denominator in Eq. (\ref{A1}) equals the collision rate $I$,
just shown to be finite for $r>r_c$ and infinite for $r<r_c$. In
the latter case, we use the regularized average
\begin{equation}
\overline{\thinspace |v|^q}=\lim_{\lambda\to
0}{\int_\lambda^\infty dv\thinspace
v^{q+1}P(0,v)\over\int_\lambda^\infty dv\thinspace
vP(0,v)}\thinspace,\quad r<r_c\thinspace.\label{A2}
\end{equation}
From Eqs. (\ref{A1}), (\ref{A2}), and the asymptotic form
(\ref{asy1}), (\ref{asy2}) of $P(0,v)$ for small $v$, one sees
that all the moments $\overline{\thinspace |v|^q}$ with $q>0$
collapse at $r=r_c$. For $r>r_c$ they are finite and nonzero, and
for $r< r_c$ they vanish.

CSB \cite{csb} analyzed the case of a randomly accelerated
particle, initially at $x=0$ with $v_0>0$, moving on the half line
$x>0$ with inelastic collisions at $x=0$. Defining $ Q_n(v,v_0)dv$
as the probability of a velocity just after the $n$th reflection
between $v$ and $v+dv$, normalized so that
\begin{equation}
\int_0^\infty dv\thinspace Q_n(v,v_0)=1\thinspace,\label{normQ}
\end{equation}
they calculated $Q_n(v,v_0)$ and the moments
\begin{equation}
\overline{\thinspace|v_n|^q}=\int_0^\infty dv\thinspace v^q\
Q_n(v,v_0)\label{moments2}
\end{equation}
exactly. In the limit $n\to\infty$, the $q$th moment diverges,
independent of $r$, for $q>{1\over 2}$. For $0<q<{1\over 2}$, this
same quantity diverges for $r>r^*(q)$ and vanishes for $r<r^*(q)$.
The critical parameter $r^*(q)$, given by Eq. (\ref{asy2}) with
the replacement $\beta\to q+2$, decreases monotonically from $r_c$
to $0$ as $q$ increases from $0$ to ${1\over 2}$. Thus, in both
the semi-infinite geometry $x>0$ and the finite geometry $0<x<1$,
certain moments of the reflected velocity collapse as $r$
decreases. However, since boundary collisions are less frequent in
the semi-infinite geometry, the velocity fluctuations are greater,
and the collapse is less complete. In the semi-infinite case the
uncollapsed moments are infinite, the moments with $q>{1\over 2}$
do not collapse, and for $0<q<{1\over 2}$ the critical parameter
$r^*(q)$ is less than $r_c$.

For a particle confined to $x<0<1$ rather than $x>0$, the
recurrence relation that determines $Q_n(v,v_0)$ is given by
\begin{eqnarray}
&&rQ_{n+1}(rv,v_0)=\int_0^\infty du\thinspace v G(1,v,u)Q_n(u,v_0)\thinspace,\label{recurr1}\\
&&Q_0(v,v_0)=\delta(v-v_0)\thinspace,\label{recurr2}
\end{eqnarray}
as shown in the Appendix. The kernel $G(1,v,u)$ is the same as in
Eqs. (\ref{ieq1}) and (\ref{ieq2}). Due to the property
(\ref{normG}) (see Appendix) of the kernel, the recurrence
relation preserves the normalization (\ref{normQ}). In the limit
$n\to\infty$, Eq. (\ref{recurr1}) becomes identical with the
integral equation (\ref{ieq1}) for $vP(0,v)$, suggesting that
$Q_\infty(v,v_0)$ is proportional to $vP(0,v)$.

This proportionality could have been anticipated. In the limit
$n\to \infty$, $Q_n(v,v_0)$ is expected to approach the
equilibrium distribution $Q_{\rm equil}(v)$, and $Q_{\rm
equil}(v)$ proportional to $vP(0,v)$ follows from the
interpretation of $vP(0,v)dv$ as the reflected probability
current, in equilibrium, in the range $v$ to $v+dv$.

The proportionality constant is fixed by the normalization
(\ref{normQ}). This leads to
\begin{equation}
Q_{\rm equil}(v)={vP(0,v)\over \int_0^\infty dv\thinspace
vP(0,v)}\thinspace,\quad r>r_c\thinspace.\label{Qequil1}
\end{equation}
For $r<r_c$ the denominator in Eq. (\ref{Qequil1}), which equals
the collision rate $I$, diverges. Regularizing as in Eq.
(\ref{A2}), we replace the right side of Eq. (\ref{Qequil1}) by
$\lim_{\lambda\to 0}Q(v,\lambda)$, where
\begin{equation}
Q(v,\lambda)={\theta(v-\lambda)vP(0,v)\over \int_\lambda^\infty
dv\thinspace vP(0,v)}\thinspace,\label{Qequil2}
\end{equation}
and $\theta(x)$ denotes the standard step function. Since
$\int_0^\infty dv\thinspace Q(v,\lambda)=1$, and since
$Q(v,\lambda)$ vanishes in the limit $\lambda\to 0$ except at
$v=0+$, where it diverges,
\begin{equation}
Q_{\rm equil}(v)=\lim_{\lambda\to
0}Q(v,\lambda)=\delta(v)\thinspace,\quad
r<r_c\thinspace.\label{Qequil3}
\end{equation}
The distribution function $Q_{\rm equil}(v)$ collapses from
(\ref{Qequil1}) to (\ref{Qequil3}) as $r$ is lowered through
$r_c$. The vanishing of the moments $\overline{\thinspace
|v|^q}=0$, $q>0$ for $r<r_c$ is consistent with the collapsed form
(\ref{Qequil3}).

That $Q_{\rm equil}(v)$ in Eq. (\ref{Qequil1}) is indeed a
stationary solution of the recurrence relation (\ref{recurr1})
follows directly from the integral equation (\ref{ieq1}) satisfied
by $P(0,v)$. That the delta function (\ref{Qequil3}) is a
stationary solution for any $r$ may be shown by substituting
$\delta(u-\epsilon)$, $\epsilon>0$ on the right side of Eq.
(\ref{recurr1}), integrating over $u$, and then taking the limit
$\epsilon\to 0$.

\section{Closing remarks}
\subsection{Is there inelastic collapse?}
The paper of CSB \cite{csb} on inelastic collapse is almost
entirely concerned with establishing that on the half line $x>0$
(i) the particle makes an infinite sequence of boundary collisions
in a finite time for $r<r_c$, and (ii) in the limit $n\to\infty$
the reflected velocity distribution $Q_n(v,v_0)$ and certain
moments of the reflected velocity collapse as $r$ is lowered. Our
results for a particle in equilibrium on the finite line $0<x<1$
are quite compatible with (i) and (ii). We question only the
statement, below Eq. (19) of \cite{csb}, that after undergoing an
infinite sequence of collisions the particle {\em remains at rest
on the boundary}.

Unlike the central quantity $Q_n(v,v_0)$ in the work of CSB, the
solution $P(x,v)$ of the Fokker-Planck equation
(\ref{fp})-(\ref{probcon}) provides information on both the
position and velocity of the particle in equilibrium. The solution
$P(x,v)$ that we have obtained does not collapse onto the
boundaries $x=0$ and $x=1$ as $r$ is lowered between $1$ and $0$.
However, for $r<r_c$, $P(0,v)$ diverges more strongly than
$v^{-2}$ in the limit $v\to 0$, and this implies $I=\infty$,
$\overline{\thinspace |v|^q}=0$, $q>0$, and $Q_{\rm
equil}(v)=\delta(v)$, via Eqs. (\ref{collrate}), (\ref{A2}), and
(\ref{Qequil3}). There is a collapse transition in the
distribution of reflected velocities $Q_{\rm equil}(v)$, but it
does not involve localization of the particle at the boundaries.

Why is $Q_{\rm equil}(v)=\delta(v)$ not a sufficient condition for
inelastic collapse? Since the velocity $v=0$ on reflection from
the boundary is overwhelmingly favored, doesn't the particle
remain at the boundary? In our opinion the relevant quantity in
the question of localization is not $Q_{\rm equil}(v)$ but the
probability per unit time $vP(0,v)dv$ for leaving the boundary
with a velocity between $v$ and $v+dv$, where
\begin{equation}
vP(0,v)=IQ_{\rm equil}(v)\thinspace,\label{answer}
\end{equation}
as in Eqs. (\ref{Qequil1}), (\ref{Qequil2}). If $vP(0,v)>0$ for
$v>0$, the particle does not remain at the boundary.

For $r<r_c$, the collision rate $I$ is infinite, and for $v>0$ the
product $IQ_{\rm equil}(v)=I\delta(v)$ on the right side of Eq.
(\ref{answer}) is indeterminate. Whether or not $vP(0,v)$ vanishes
for $v>0$ is unclear from Eq. (\ref{answer}). We have calculated
$P(0,v)$ for $r<r_c$ by solving the Fokker-Planck equation. The
result, as described above, is a smooth function of $v$ with the
asymptotic forms (\ref{asy1})-(\ref{asy3}). The quantity $vP(0,v)$
does not vanish for $v>0$, although it does indeed imply $Q_{\rm
equil}(v)=\delta(v)$. Thus, we find that inelastic collisions do
not localize the particle at the boundaries.

Below we comment on two earlier results in view of these
conclusions.

\subsection{Collision rate in simulations}
In computer simulations \cite{fbb,la,bfg}with a discrete time step
$\Delta t$, the boundary collision rate $I$, which can never
exceed one collision per time step, is necessarily finite. In the
algorithm of \cite{la,bfg}, the root-mean-square velocity change
is given by $\Delta v=(2\Delta t)^{1/2}$. In the limit $\Delta
t\to 0$ the discrete dynamics approaches the continuum dynamics of
Eq. (\ref{eqmo}), and $I$ diverges for $r\leq r_c$. Anton
\cite{la} has found that the collision rate in his simulations
scales as $I\sim(\Delta t)^{(2-\beta)/2}$, $\Delta t\to 0$ for
$r<r_c$ and offered a dynamical explanation. We note that this
scaling relation follows very simply from our results for the
equilibrium distribution function $P(x,v)$. For velocities
$|v|{\stackrel{<}{\sim}}\Delta v$, the simulation results are
expected to deviate from the asymptotic form $P(0,v)\sim
v^{-\beta(r)}$ in Eqs. (\ref{asy1}), (\ref{asy2}). Thus the
boundary collision rate (\ref{collrate}) in the simulations scales
as
\begin{equation}
I\sim\int_{\Delta v}^\infty dv\thinspace v^{1-\beta}\sim (\Delta
v)^{2-\beta}\sim (\Delta t)^{(2-\beta)/2}\thinspace,\quad \Delta
t\to 0\thinspace.\label{coll2}
\end{equation}

\subsection{Persistence Exponent for ${\bf r<r_c}$}

Burkhardt \cite{twb2} and De Smedt {\it et al.} \cite{dsgl} have
considered the probability $Q(x_0,v_0,t)$ that a randomly
accelerated particle with initial position and velocity $x_0$,
$v_0$, confined to the half-line $x>0$ and reflected inelastically
at $x=0$, has not yet {\em undergone inelastic collapse} after a
time $t$. They predicted $Q(x_0,v_0,t)=1$ for $r>r_c$, and for
$r<r_c$ the power-law decay
\begin{equation}
Q(x_0,v_0;t)\sim t^{(2-\beta)/2}\thinspace,\quad
t\to\infty\thinspace,\label{survprob}
\end{equation}
where the exponent $\beta$ is the same as in Eqs. (\ref{asy1}),
(\ref{asy2}), and (\ref{coll2}). In view of our conclusions that
the particle makes an infinite number of collisions in a finite
time but does not remain at the boundary, $Q(x_0,v_0,t)$ in Eq.
(\ref{survprob}) should be interpreted as the probability that
after a time $t$ the randomly-accelerated particle has not yet
made an infinite number of boundary collisions. The derivations of
Eq. (\ref{survprob}) in \cite{twb2,dsgl} are compatible with this
interpretation, and it is also supported by simulations
\cite{la,sb,kb}.

\acknowledgements We thank Jerrold Franklin for numerical
calculations of the curves in Figs. 1 and 2 and for many
stimulating discussions. TWB also gratefully acknowledges
discussions and correspondence with Lucian Anton and Alan Bray.

\appendix
\section{Velocity distribution on arrival at the boundary}

The probability that a randomly accelerated particle with initial
position $x=0$ and initial velocity $u>0$, moving on the half line
$x>0$, arrives with speed between $v$ and $v+dv$ on its first
return to $x=0$ is given by $vG_0(v,u)dv$, where
\begin{equation}
G_0(v,u)= {3\over 2\pi}\thinspace{v^{1/2}u^{1/2}\over
v^3+u^3}\thinspace.\label{G0}
\end{equation}
This result, due to McKean \cite{mck}, was also obtained
independently by CSB \cite{csb}.

The quantity $G(1,v,u)$ in Eq. (\ref{ieq2}), derived by BFG
\cite{bfg}, extends this result to the the finite interval
$0<x<1$. The probability that a randomly accelerated particle
which leaves $x=0$ with velocity $u>0$ has speed between $v$ and
$v+dv$ the next time it reaches either boundary is given by
$vG(1,v,u)dv$, where the first and second terms on the right side
of Eq. (\ref{ieq2}) correspond to arrival at $x=0$ and $x=1$,
respectively. Like $G_0(v,u)$ in Eq. (\ref{G0}), $G(1,v,u)$
satisfies the normalization condition
\begin{equation}
\int_0^\infty dv\thinspace vG(1,v,u)=1\thinspace.\label{normG}
\end{equation}

Integral equation (\ref{ieq1}) for $P(0,v)$ follows directly from
the interpretation of $G(1,v,u)$ in the preceding paragraph and
the stationarity, in equilibrium, of the reflected current
$vP(0,v)dv$ between $v$ and $v+dv$. Another consequence is the
recurrence relation (\ref{recurr1}) for the probability
distribution $Q_n(v,v_0)$ of the speed with which the particle
rebounds after the $n$th boundary collision. Solving Eqs.
(\ref{recurr1}), (\ref{recurr2}) with $G_0(v,u)$ in Eq. (\ref{G0})
in place of $G(1,v,u)$, CSB \cite{csb} calculated $Q_n(v,v_0)$
exactly for motion on the half line $x>0$.

\begin{figure}[f1a]
\begin{center}
\includegraphics*[width=0.7\textwidth]{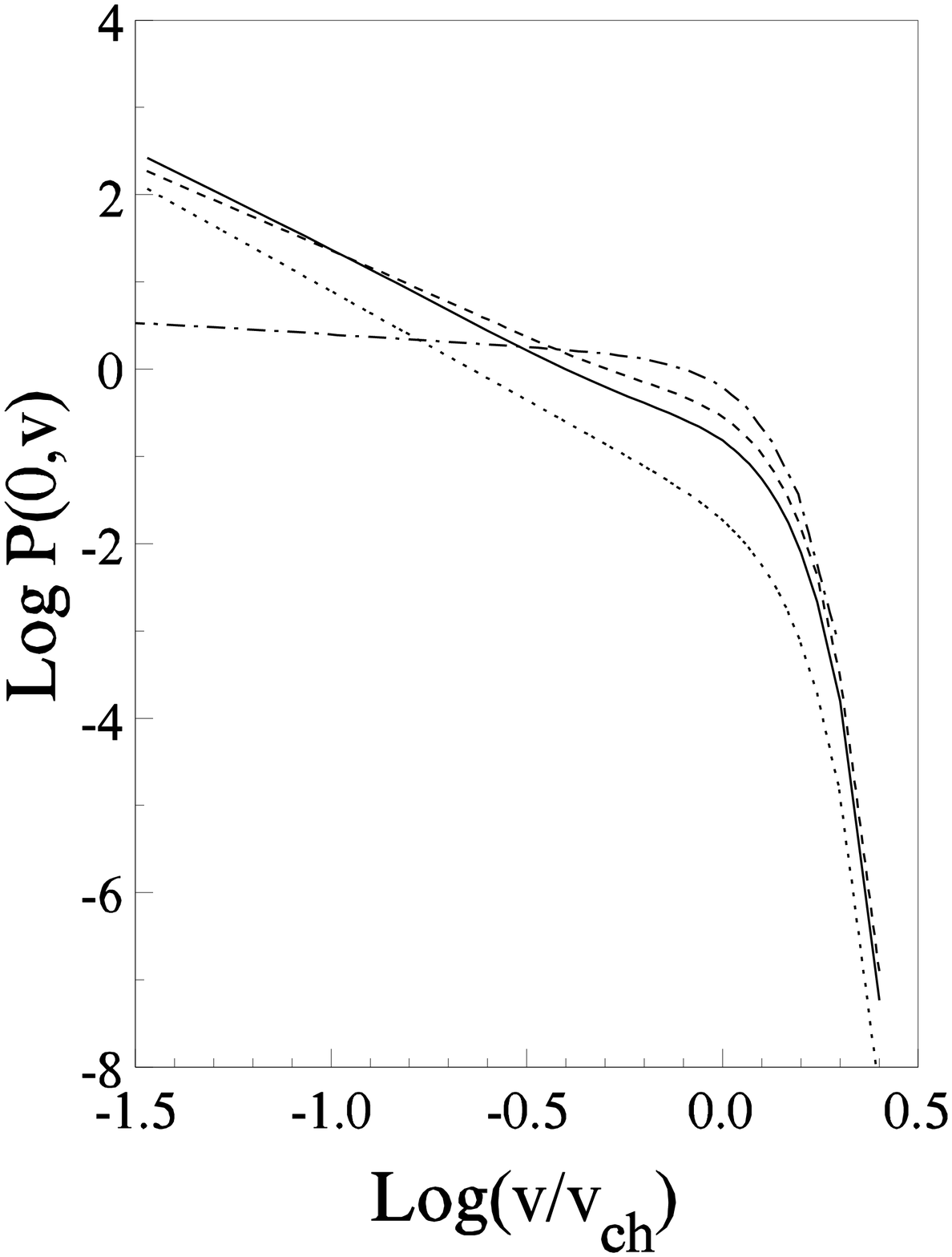}\\*
\end{center}
\caption{Double-logarithmic plot (base 10) of $P(0,v)$ for
$r=0.01$ (dotted curve), $r=0.1$ (solid curve), $r=0.2$ (dashed
curve), and $r=0.5$ (dot-dashed curve). The curves are normalized
according to Eq. (\ref{normP}).} \label{fig1}
\end{figure}
\newpage

\begin{figure}[f2a]
\begin{center}
   \includegraphics*[width=0.7\textwidth]{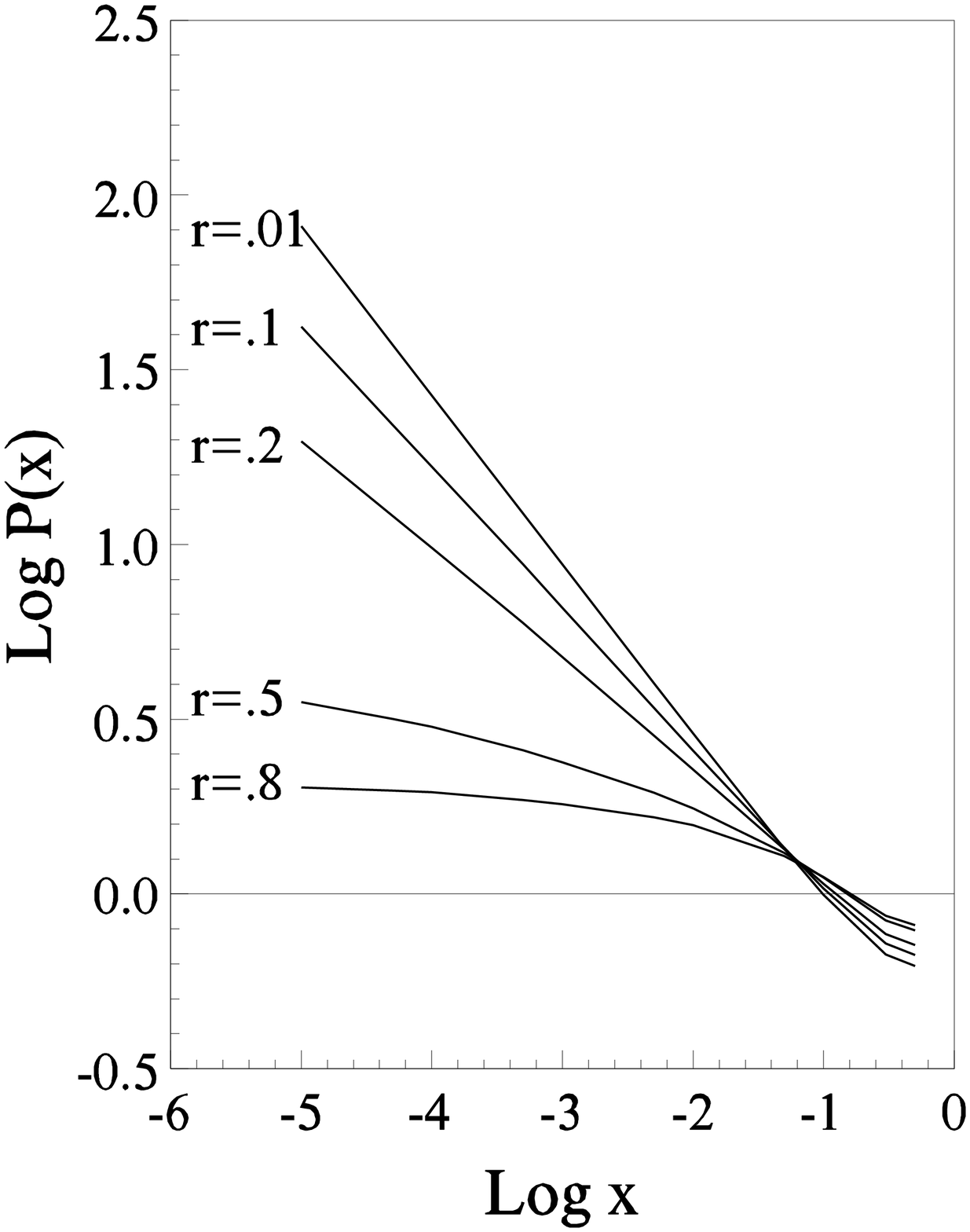}\\*
\end{center}
\caption{Double-logarithmic plot (base 10) of $P(x)$ for several
values of $r$. The curves are normalized according to Eq.
(\ref{normP}).} \label{fig2}
\end{figure}
\newpage


\newpage
\begin{thebibliography}{03}
\bibitem{csb} S. J. Cornell, M. R. Swift, A. J. Bray, Phys. Rev.
Lett. {\bf 81}, 1142 (1998).
\bibitem{fbb} J. Florencio, F. C. S\'a Barreto, and O. F. de Alcantura
Bonfim, Phys. Rev. Lett. {\bf 84}, 196 (2000).
\bibitem{la} L. Anton, Phys. Rev. E {\bf 65}, 047102 (2002).
\bibitem{bfg} T. W. Burkhardt, J. Franklin, and R. R. Gawronski,
Phys. Rev. E {\bf 61}, 2376 (2000).
\bibitem{overline} We use an overline to denote averages weighted with the boundary collision
rate $vP(0,v)dv$ in the range $v$ to $v+dv$, as in Eq. (\ref{A1}),
and reserve angular brackets for equilibrium averages of the form
$\langle A\rangle=\int_0^1 dx\int_{-\infty}^\infty dv\thinspace
A(x,v)P(x,v)$.
\bibitem{mp} J. Masoliver and J. M. Porr\`a, Phys. Rev. Lett. {\bf
75}, 189 (1995); Phys. Rev. E {\bf 53}, 2243 (1996).
\bibitem{gr} I. S. Gradshteyn and I. M. Ryzhik, {\it Tables of
Integrals, Series, and Products} (Academic, New York, 1980).
\bibitem{mck} H. P. McKean, J. Math. Kyoto Univ. {\bf
2}, 227 (1963).
\bibitem{smallu} For small $u$, $G(x,v,u)\approx{1\over
6}\pi^{-3/2}u^{1/2}v^2x^{-3/2}(1-x)^{-3/2}\exp[-(1-2x)w][(1-2x)K_{1/3}(w)
+K_{2/3}(w)]$, where $w=v^3/[18x(1-x)]$.
\bibitem{twb2} T. W. Burkhardt J. Phys. A {\bf 33}, L429
(2000); Phys. Rev. E {\bf 63}, 011111 (2001).
\bibitem{dsgl} G. De Smedt, C. Godr\`eche, and J. M. Luck, Europhys. Lett.
{\bf 53}, 438 (2001).
\bibitem{sb} M. R. Swift and A. J. Bray, Phys. Rev. E {\bf
59}, R4721 (1999).
\bibitem{kb} S. N. Kotsev and T. W. Burkhardt, to be published.
\end{thebibliography}
\end{document}